\documentclass[conference]{IEEEtran}
\usepackage{color}
\usepackage{multirow}
\usepackage{cite}
\usepackage{balance}
\usepackage{amsmath}
\usepackage{amssymb}
\usepackage{enumitem}
\usepackage{arydshln}
\usepackage{fancyhdr}
\usepackage{marvosym}
\setlist[itemize]{noitemsep, topsep=0pt}
\usepackage{pdfsync}
\ifCLASSINFOpdf
\usepackage[pdftex]{graphicx}
\DeclareGraphicsExtensions{.pdf,.jpeg,.png}
\else
\usepackage[dvips]{graphicx}
\DeclareGraphicsExtensions{.eps}
\fi
\graphicspath{{figs/}}
\usepackage{epstopdf}
\usepackage{booktabs}
\usepackage{amsmath}

\renewcommand{\footnoterule}{%
  \kern -3pt
  \hrule width 0.49 \textwidth height 0.5pt
  \kern 1pt
}
\usepackage[absolute]{textpos}
\TPMargin{5pt}

\newcommand{\copyrightstatement}{
    \begin{textblock}{0.5}(0.08,0.94)   
    \noindent
         \footnotesize
          978-1-7281-0407-2/19/\$31.00~\copyright2019 IEEE 
    \end{textblock}
}

\begin{document}

\copyrightstatement
	
	\title{Distributed PV Penetration Impact Analysis on Transmission System Voltages using Co-Simulation}
	
	
	\author{\IEEEauthorblockN{Gayathri Krishnamoorthy, R. Sadnan, and Anamika Dubey}
		\IEEEauthorblockA{\textit{School of Electrical Engineering and Computer Science} \\
			\textit{Washington State University} \\
			\textit{Pullman, WA}\\
		}
}
	\maketitle
	
	\begin{abstract}
		With the growing penetrations of distributed energy resources (DERs), it is imperative to evaluate their impacts on transmission system operations. In this paper, an iteratively coupled transmission and distribution (T\&D) co-simulation framework is employed to study the impacts of increasing penetrations of distribution-connected photovoltaic (PV) systems on transmission system voltages. The co-simulation framework introduces iterative coupling and unbalanced transmission system analysis that help accurately replicate the standalone T\&D system results without resorting to the computational challenge of developing large-scale standalone T\&D models. The integrated T\&D systems are evaluated for multiple PV deployment scenarios based on randomly generated PV locations and sizes. A test system is simulated using IEEE-9 bus transmission system model integrated at each load point with three EPRI's Ckt-24 distribution feeder models. The results are thoroughly validated using a standalone T\&D system model simulated in OpenDSS. 	
	\end{abstract}
	\begin{IEEEkeywords}
		Transmission \& Distribution co-simulation, photovoltaics (PV), distributed energy resources (DERs).
	\end{IEEEkeywords}
	
	\IEEEpeerreviewmaketitle
	
	\section{Introduction}
	
	With the incentivized rapid decarbonization of electric power generation industry and aggressive renewable portfolio standards (RPS) in most states, the electric power delivery (T\&D) system is rapidly transforming into a decentralized, bidirectional network \cite{curtright2008character,29680,nelsen2009megawatts}. For instance, participation of distributed energy resources (DERs) has increased its levels where in 2017, 6.3\% of the total U.S. demand was supplied by wind turbine technology (WT) and 1.83\% by distributed and utility-scale photovoltaic units (PVs) \cite{Data}. Several exploratory studies and field demonstrations have pointed out that the integration of DERs is increasing the stress on power delivery systems; high penetration of distribution-level solar photovoltaics (PV) are known to cause over voltage problems, reverse power flow, increased power losses and power quality issues \cite{eber2013hawaii,dubey2016analytical,hawkins2007integration, smith2012stochastic, dubey2015understanding}. For this reason, utility companies/distribution system operators perform DER interconnection study prior to permitting a new DER/PV connection request.
	
	Unfortunately, most of the existing DER interconnection studies perform DER impacts analysis either only at the distribution level or individually at transmission and distribution level using a decoupled T\&D system model\cite{evans2014new}. The potential impacts on the transmission grid are either ignored given low penetrations of DERs, or are non-representative due to the decoupled T\&D model. With growing penetrations, the distribution-level DERs are expected to affect both transmission and distribution systems \cite{palmintier2016integrated}. Thus, DER impact assessment requires a very detailed study that not only needs to be performed both at transmission and distribution levels but also should evaluate the interactions between T\&D systems during high DER penetrations. The decoupled analysis of T\&D system is no longer adequate, calling for new tools capable of capturing the interactions between the transmission and distribution systems.
	 
	 Recently, this led to the development of multiple T\&D co-simulation platforms. The existing co-simulation platforms for integrated T\&D system analysis, however, use a balanced positive sequence AC power flow for transmission system analysis and loosely couple T\&D networks \cite{ciraci2014fncs, palmintier2017igms, Helics}. With increasing levels of system unbalance in the distribution system due to single-phase small-scale DERs, a balanced positive sequence approach may not accurately model integrated T\&D systems. Furthermore, a loosely coupled co-simulation approach may not accurately model T\&D operations with faster changes in DER generation patterns. Also, most of the existing T\&D co-simulation framework have not been validated against an equivalent standalone T\&D model \cite{huang2017integrated, huang2017comparative}.
	
	To address these concerns, in our prior work, we proposed an iterative coupled T\&D co-simulation framework that also supports three-phase transmission system analysis. Specifically, we model transmission system in three-sequence detail to accurately capture the system unbalance and develop a fixed-point iteration (FPI) technique to tightly couple the simulators (T\&D) using an iterative approach that ensures convergence at the point of common coupling (PCC) \cite{DubeyPES2018}. In this paper we use our previously developed co-simulation framework to evaluate the voltage impacts of increasing distributed PV penetrations in distribution grid. We also thoroughly validate the co-simulation results against the standalone model developed for the same test system using OpenDSS, a commercially available distribution system simulator \cite{OpenDSS}. The main contributions of this paper are as follows.
	
	\begin{itemize}[nolistsep,leftmargin=*]
		\item A detailed analysis of convergence characteristics of the proposed iterative framework is presented by increasing the PV penetrations to simulate varying levels of unbalance in the distribution system loading.
		\item The impacts of increasing levels of PV penetration on transmission system voltages and real power are analyzed with the help of co-simulation approach. 
		\item The observed co-simulation results are thoroughly validated by simulating similar conditions with PVs in the OpenDSS standalone T\&D model.
	\end{itemize}


	\section{Integrated T\&D Modeling Framework}
	
	The primary objective of this paper is to use the developed iterative T\&D co-simulation framework to understand how the distribution connected PVs impact the normal transmission system operations. In this study, the transmission system is developed in MATLAB and includes a detailed three-sequence network model. The unbalanced distribution system analysis is preformed using OpenDSS--commonly used distribution system modeling and analysis software. The transmission and distribution  models, solved individually using MATLAB and OpenDSS, respectively, are iteratively coupled using a co-simulation script developed in MATLAB. Please refer to \cite{DubeyPES2018} for details on co-simulation framework.
	
	\subsection{Transmission and Distribution System Quasi-static model}
	
	A detailed three-phase power flow analysis is required for transmission system analysis to accurately capture the impacts of distribution-introduced demand unbalance. In this study, a three-sequence model for transmission system is developed using sequence component method \cite{zhang1994asymmetrical,smith1998improved,abdel2005improved}. The three-sequence model can represent the effects of unbalanced loads and untransposed transmission lines while not significantly increasing the computational complexity of the power flow analysis. The three-phase transmission system is decoupled into three independent sequence circuits by replacing the off-diagonal elements with the respective compensation current injections \cite{abdel2005improved}. The decoupled three-sequence models are solved separately. The positive sequence model is solved using Newton-Raphson method. The negative and zero sequence models are solved using linear equations. The process is iterative and repeated until the change in positive sequence power flow due to negative and zero sequence components is within the pre-specified tolerance. Please refer to \cite{abdel2005improved} for details. Similarly, distribution system is modeled in full three-phase representation. The three-phase modeling and quasi-static analysis is done using OpenDSS \cite{OpenDSS}.
	
	\begin{figure}[ht]
		\centering
		\includegraphics[width=0.45\textwidth]{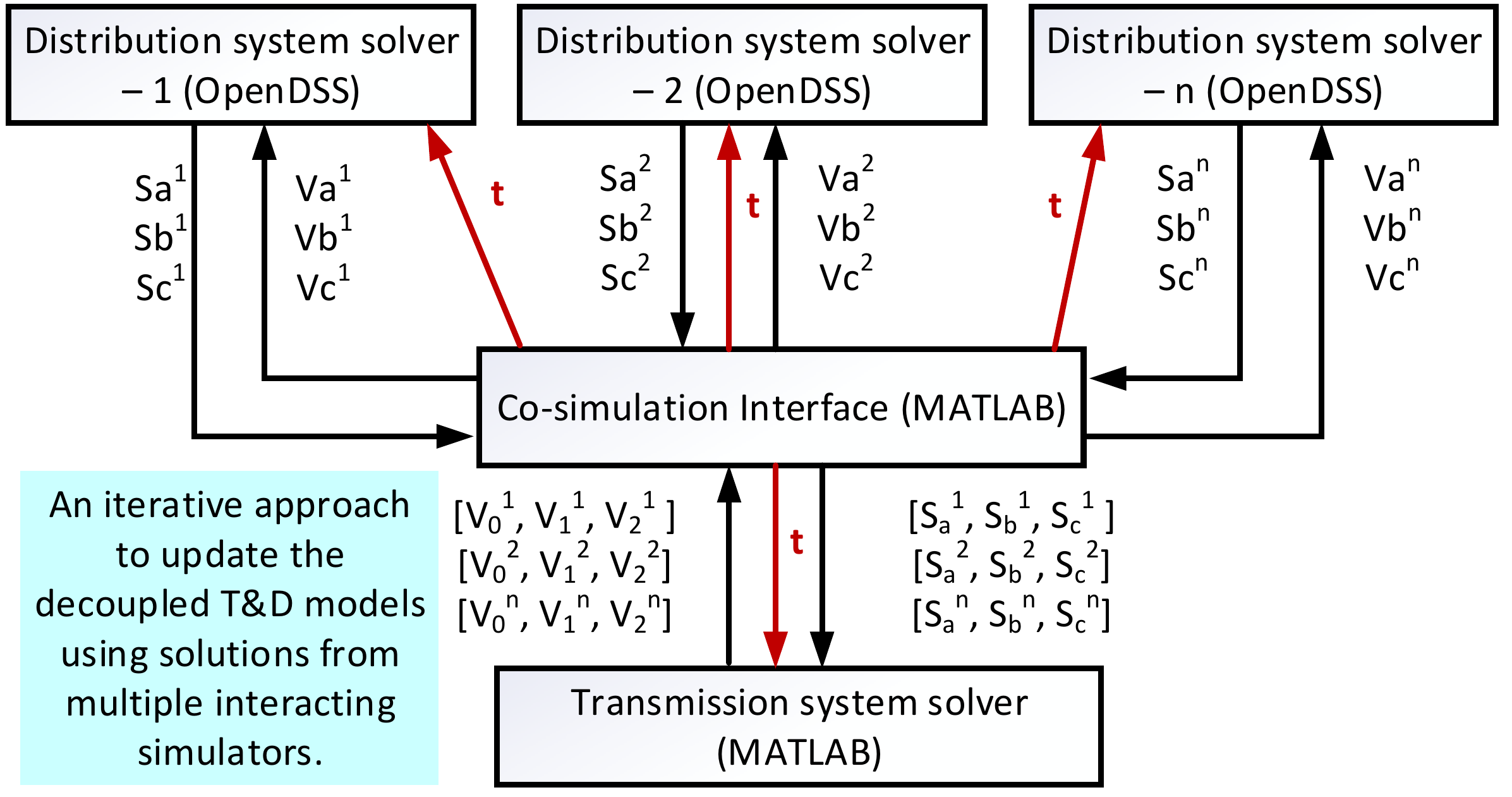}
		\vspace{-0.1cm}
		\caption{Iterative framework for the co-simulation approach at PCC}
		\label{fig:1}
		\vspace{-5 pt}
	\end{figure}
	
	The general principle of the proposed co-simulation approach is detailed here. At a given time step of quasi-static simulation, the T\&D systems are solved separately using their respective solvers. This step solves decoupled T\&D models. The transmission system solver models the connected distribution network as an equivalent load and the distribution solver models the upstream transmission bus as a voltage source. The bus voltages and angles obtained from transmission network solver and active and reactive power flow obtained from distribution network solver at the point-of-common coupling (PCC) are referred to as boundary variables (see Fig. 1). After decoupled T\&D systems are solved, boundary variables are tested for convergence. If tolerance limit is not satisfied, the co-simulation stage begins. At a given co-simulation iteration, T\&D systems have partially correct information about the boundary variables that needs updating. The decoupled representations for T\&D networks are updated at each co-simulation iteration. With updated boundary variables, the decoupled T\&D models are solved again. The co-simulation iterations are repeated until the error in boundary variables obtained from decoupled models is within the pre-specified tolerance. This leads to a co-simulation model that closely approximates a stand-alone unified model for the two systems.
	
	\subsection{PV Deployment Scenarios}
	High-levels of distribution-connected PV penetration may result in multiple operational challenges for the integrated T\&D systems including but not limited to over voltages, excessive reverse power flow, increased power losses, severe phase unbalances, and power quality issues. This study specifically focuses on analyzing the impacts of distribution-connected PVs on transmission system voltages. In this section, we detail the method used to generate random PV deployment scenarios for a given distribution feeder. Following the related literature concerning PV hosting analysis for distribution feeders, similar stochastic analysis framework is adapted to generate numerous PV deployment scenarios \cite{dubey2017estimation}. Multiple scenarios are simulated to fully capture the randomness associated with PV sizes and deployment locations for increasing customer penetration levels.
		
	The method to simulate stochastic PV deployment scenarios is briefly detailed here. For each customer penetration level, multiple unique PV deployment scenarios are simulated using Monte Carlo approach by associating a uniform random distribution to the PVs location and size. The hourly PV generation for each scenario is obtained from PV generation profile given in \cite{dubey2017estimation}. 100 different scenarios are created for each level of PV penetration. For each scenario, 10\% of the customers in Ckt-24 feeder are selected at random using a uniform distribution and PVs are deployed at their locations. The capacity of the installed PV is determined based on the peak load demand of the feeder and the type (residential or commercial) of the customer. The customer penetration is then increased in steps of 10\% and PV generation is obtained for each level. Please refer to \cite{dubey2017estimation} for further details on PV deployment.
	
	\begin{figure}[ht]
		\centering
		\includegraphics[width=0.45\textwidth]{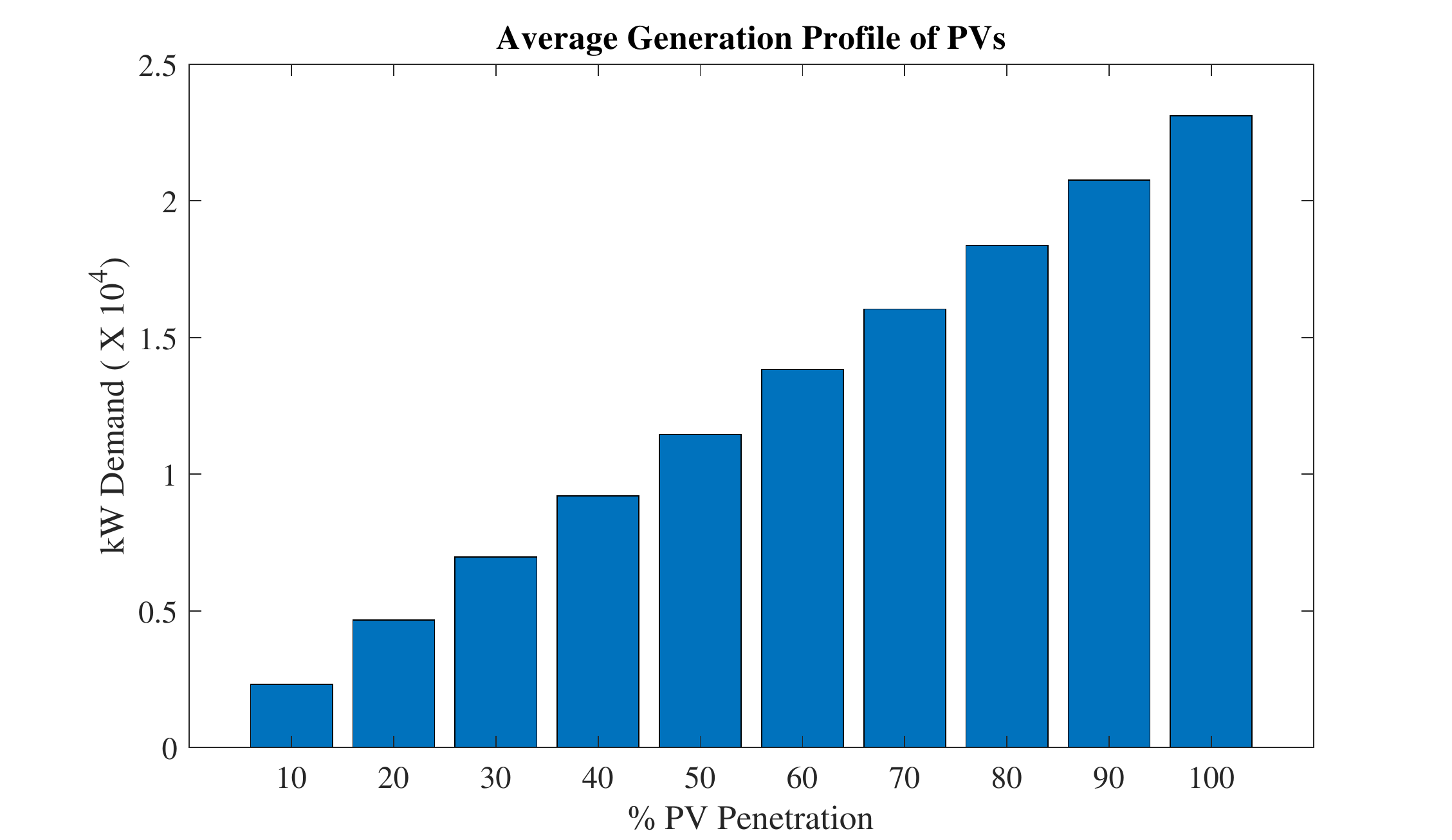}
		\vspace{-6 pt}
		\caption{Average PV generation profile}
		\label{fig:2}
		\vspace{-15 pt}
	\end{figure}
	
	The specific details of simulated PV scenarios for the distribution feeder under consideration is detailed here. EPRI Ckt-24, used as the distribution system model in this study, supplies for a total of 3885 customers at 34.5 kV voltage level using two equally loaded feeders. The peak demand recorded at the substation is 52.1 MW and 11.7 MVAR and the feeder is comprised of 87\% residential loads that includes a large number of single-phase and three-phase customers. A substation transformer is used to connect the 34.5 kV distribution system to the 230 kV transmission bus. The average PV generation profile for each of the penetration level in 100 scenarios is presented in Fig. 2.	
	
	\subsection{Stand-Alone Model Development}
	Another objective of this paper is to validate the co-simulation results against an equivalent stand-alone T\&D system model. In this paper we employ OpenDSS to simulate the stand-alone model for the selected T\&D test system. OpenDSS is a real world distribution feeder simulator that can also handle small transmission system simulations. Specifically, for stand-alone simulation, IEEE 9-bus system with three-phase details and three Ckt-24 distribution systems are connected to its load nodes is modeled in OpenDSS. The stand-alone IEEE 9-bus integrated with Ckt-24 models is solved using OpenDSS for convergence. The results obtained at the PCC are analyzed and compared with those obtained using co-simulation. In our previous work, a smaller distribution system, IEEE 13-bus distribution feeder, connected to the IEEE 9-bus system was simulated as a stand-alone model and the results were used to validate the co-simulation platform \cite{DubeyNAPS2018}. This study further validates our proposed co-simulation approach using larger integrated stand-alone T\&D system model. Since this study involves analyzing the voltage behavior of the transmission system due to increasing PV penetrations, the stand-alone model is also simulated with varying PV deployment scenarios on all Ckt-24 feeders connected to the transmission system. The voltages at the PCC obtained using co-simulation and stand-alone are compared for validation.
	
	\section{Results and Discussions}
	The test system for the analysis is described first. The transmission test system model is comprised of IEEE 9-bus test system with three generators at buses 1 (slack), 2 and 3, three loads at buses 5, 6 and 8. EPRI Ckt-24 as described in section II is used as the distribution test system \cite{TransTestFeeder,DistTestFeeder}. The integrated T\&D test system is simulated by replacing all the load points, L5, L6 and L8 of the IEEE 9-bus system by EPRI's Ckt 24 distribution feeder (see Fig 3).
	
	\vspace{-0.1cm}
	\begin{figure}[t]
		\centering
		\includegraphics[width=0.45\textwidth]{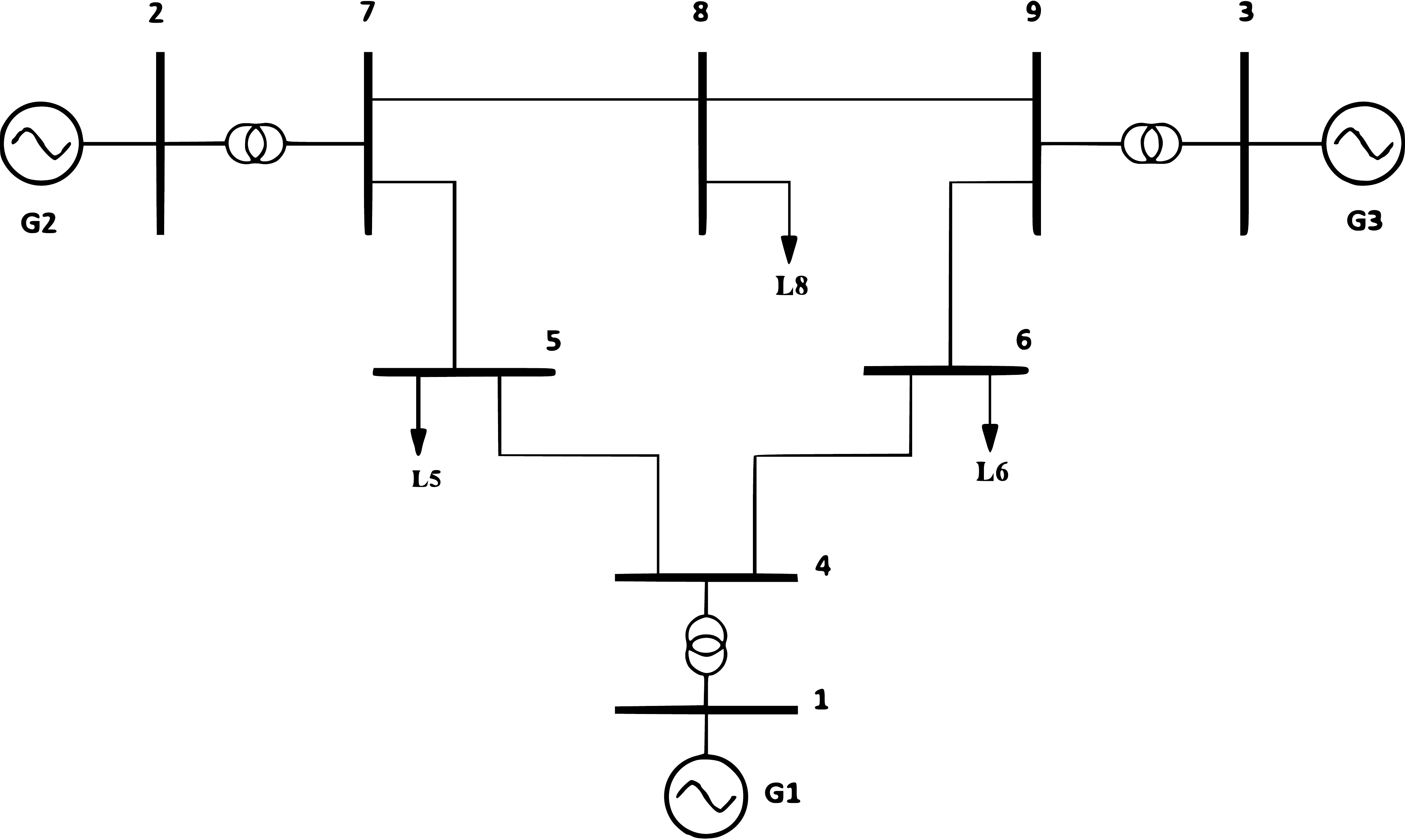}
		\vspace{-6 pt}
		\caption{Test system with multiple distribution feeders}
		\label{fig:3}
		\vspace{-50 pt}
	\end{figure}

	\subsection{Effects of Increasing Distribution-connected PVs}
	
	To capture randomness in distribution-connected PV deployment scenarios, 100 unique random scenarios each with 10 different levels of penetration are simulated. Consequently, a total of 1000 scenarios are evaluated for the given T\&D test system. 
	
		\begin{figure*}[t]
		\minipage{0.33\textwidth}
		\includegraphics[width=\linewidth, trim={0cm 0 1cm 0},clip]{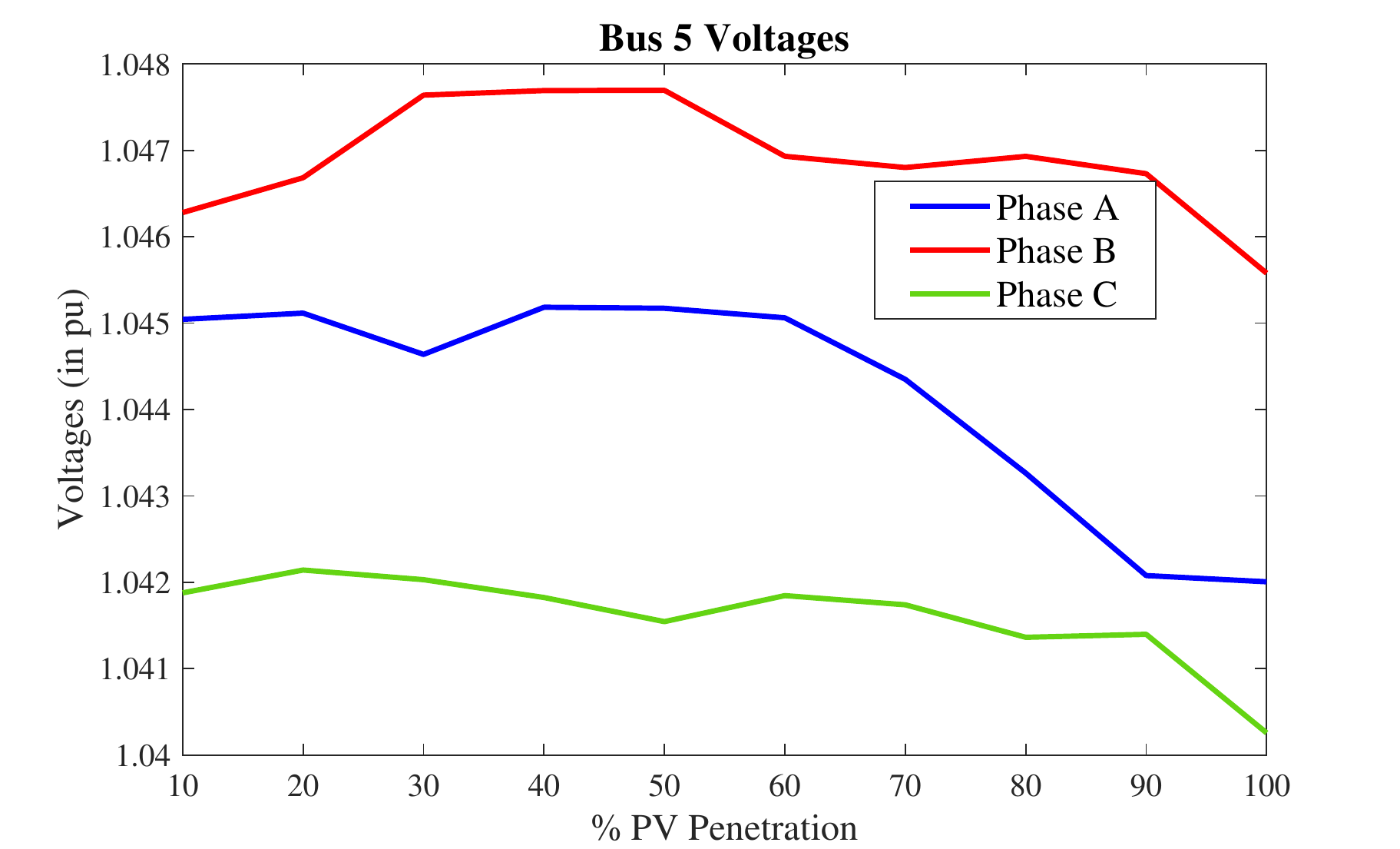}
		\endminipage\hfill
		\minipage{0.33\textwidth}
		\includegraphics[width=\linewidth,trim={0cm 0 1cm 0},clip]{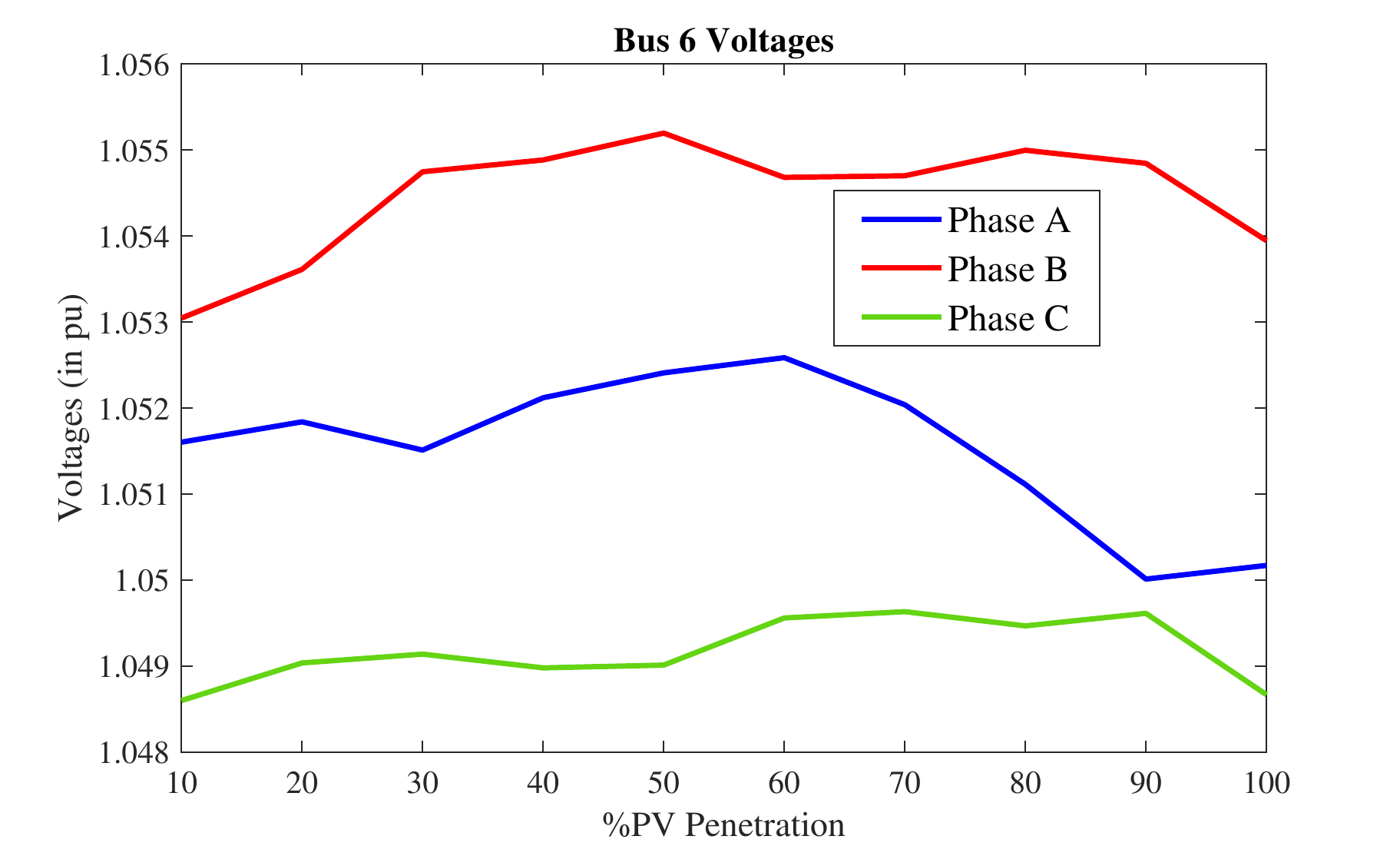}
		\endminipage\hfill
		\minipage{0.33\textwidth}%
		\includegraphics[width=\linewidth,trim={0cm 0 1cm 0},clip]{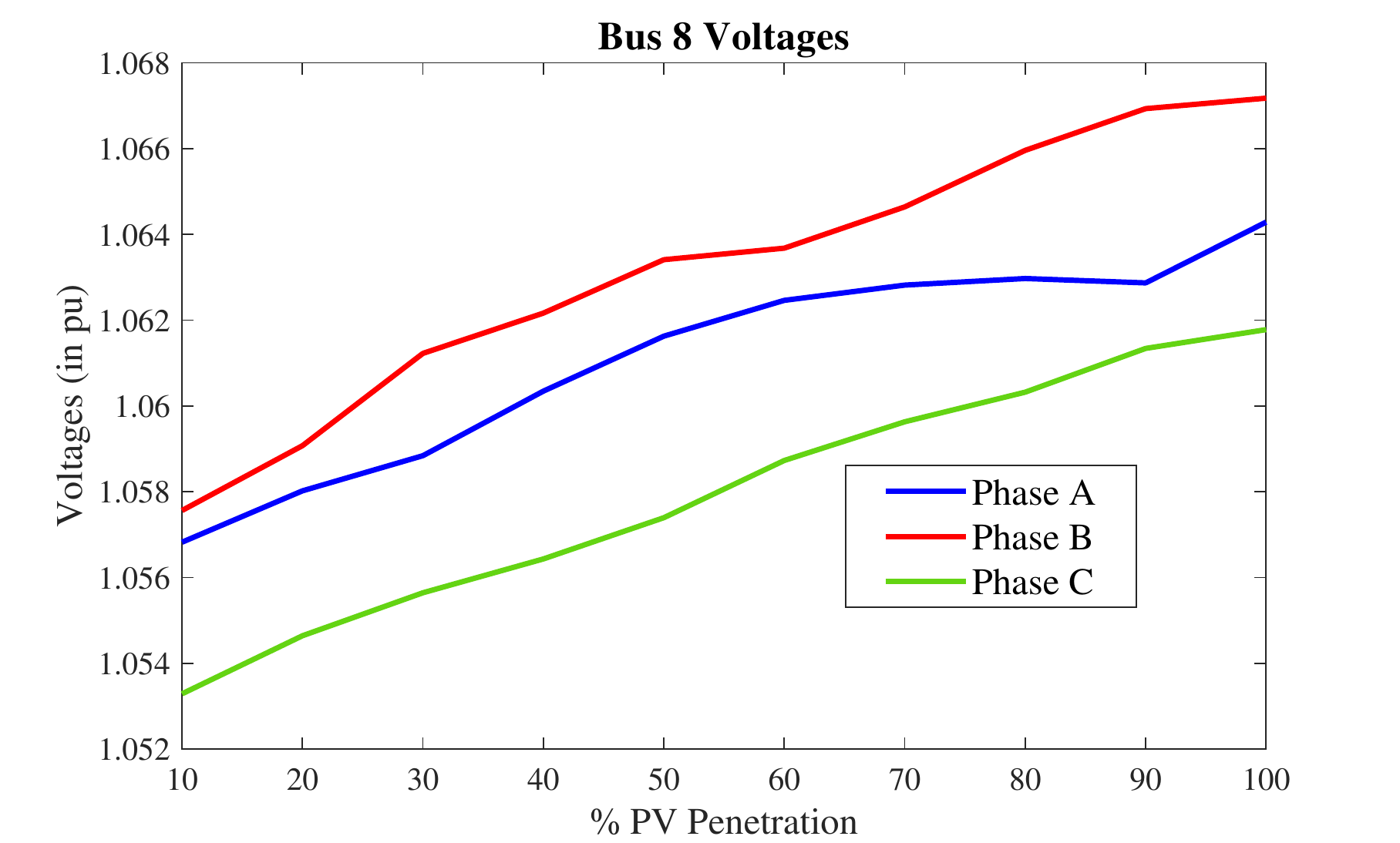}
		\endminipage
		\caption{Average of converged voltages at the PCC by varying levels of PV penetration in 100 scenarios}
		\vspace{-0.4cm}
	\end{figure*}
	
	First, the results for scenario 25 is discussed. The converged values of real power and bus voltage magnitude at PCC are presented in Table I. As seen from the Table I, with the increase in PV penetrations, the real power demand of the substations from the transmission system decreases. This effect is seen in all three PCCs. With the reduction in real power consumption by the distribution system, the voltages at the substation buses (PCC) are expected to increase. This trend is followed by PCC at bus 8, but the other two PCCs exhibit a different behavior. The voltages at the PCC of bus 6 and 5 are increasing up to 50\% penetration. When the PV penetration is increased to 60\% and further, there is a decline in the voltage magnitude at their respective PCCs. This trend was initially speculated to be a behavior of this one specific scenario. Further evaluation of these results was required. So, all 100 scenarios for each penetration level was simulated using the co-simulation platform and the average converged voltages at the PCC is studied and the results are presented in Fig 4.
	
	
	\begin{table}[t]
		\centering
		\footnotesize
		\caption{ Converged voltages and real powers at T\&D PCC for varying PV penetrations}
		\vspace{-0.2cm}
		\label{table1}
		\begin{tabular} {c|c|c|c|c|c|c|c}
			\hline
			\multirow{2}{*} {\%PV} & \multirow{2}{*} {$\Phi$}  &  \multicolumn{3}{c|}{Real Power at PCC (p.u.)}& \multicolumn{3}{c}{Voltages at PCC (p.u.)} \\
			\cline{3-8}
			& &  Bus 5 & Bus 6 & Bus 8 & Bus 5 & Bus 6 & Bus 8 \\
			\hline
			\multirow{3}{*}{ 10}	&	A	&	0.167	&	0.167	&	0.1675	&	1.045	&	1.0516	&	1.0568	\\
			&	B	&	0.1647	&	0.1648	&	0.1651	&	1.0463	&	1.053	&	1.0576	\\
			&	C	&	0.1647	&	0.1647	&	0.1651	&	1.0419	&	1.0486	&	1.0533	\\
			\hline
			\multirow{3}{*}{ 20}	&	A	&	0.1595	&	0.1596	&	0.1596	&	1.0451	&	1.0518	&	1.058	\\
			&	B	&	0.1573	&	0.1574	&	0.1573	&	1.0467	&	1.0536	&	1.0591	\\
			&	C	&	0.1571	&	0.1572	&	0.1571	&	1.0421	&	1.049	&	1.0546	\\
			\hline
			\multirow{3}{*}{ 30}	&	A	&	0.1518	&	0.1519	&	0.1515	&	1.0451	&	1.0519	&	1.0588	\\
			&	B	&	0.1496	&	0.1496	&	0.1492	&	1.0476	&	1.0547	&	1.0612	\\
			&	C	&	0.1487	&	0.1488	&	0.1483	&	1.042	&	1.0491	&	1.0556	\\
			\hline
			\multirow{3}{*}{ 40}	&	A	&	0.1439	&	0.1444	&	0.1442	&	1.0452	&	1.0521	&	1.0603	\\
			&	B	&	0.1413	&	0.1418	&	0.1414	&	1.0477	&	1.0549	&	1.0622	\\
			&	C	&	0.1409	&	0.1414	&	0.141	&	1.0418	&	1.0493	&	1.0564	\\
			\hline
			\multirow{3}{*}{ 50}	&	A	&	0.1353	&	0.1353	&	0.1356	&	1.0452	&	1.0524	&	1.0616	\\
			&	B	&	0.1324	&	0.1325	&	0.1326	&	1.0477	&	1.0552	&	1.0634	\\
			&	C	&	0.1321	&	0.1322	&	0.1323	&	1.0418	&	1.0495	&	1.0574	\\
			\hline
			\multirow{3}{*}{ 60}	&	A	&	0.1282	&	0.1278	&	0.1286	&	1.0451	&	1.0526	&	1.0625	\\
			&	B	&	0.1257	&	0.1254	&	0.126	&	1.0469	&	1.0547	&	1.0637	\\
			&	C	&	0.1256	&	0.1253	&	0.1259	&	1.0417	&	1.0496	&	1.0587	\\
			\hline
			\multirow{3}{*}{ 70}	&	A	&	0.1211	&	0.1208	&	0.1216	&	1.0443	&	1.052	&	1.0628	\\
			&	B	&	0.1189	&	0.1185	&	0.1192	&	1.0468	&	1.0547	&	1.0646	\\
			&	C	&	0.1184	&	0.1181	&	0.1188	&	1.0417	&	1.0496	&	1.0596	\\
			\hline
			\multirow{3}{*}{ 80}	&	A	&	0.1137	&	0.1134	&	0.1143	&	1.0433	&	1.0511	&	1.063	\\
			&	B	&	0.1115	&	0.1112	&	0.112	&	1.0467	&	1.054	&	1.066	\\
			&	C	&	0.1105	&	0.1102	&	0.111	&	1.0414	&	1.0495	&	1.0603	\\
			\hline
			\multirow{3}{*}{ 90}	&	A	&	0.1058	&	0.1059	&	0.1069	&	1.0421	&	1.05	&	1.0629	\\
			&	B	&	0.1042	&	0.1043	&	0.1052	&	1.0467	&	1.0538	&	1.0669	\\
			&	C	&	0.1025	&	0.1026	&	0.1034	&	1.0414	&	1.0496	&	1.0613	\\
			\hline
			\multirow{3}{*}{ 100}	&	A	&	0.0967	&	0.0968	&	0.098	&	1.042	&	1.0502	&	1.0643	\\
			&	B	&	0.0946	&	0.0948	&	0.0958	&	1.0456	&	1.0539	&	1.0672	\\
			&	C	&	0.0937	&	0.0938	&	0.0948	&	1.0403	&	1.0487	&	1.0618	\\
			
			\hline	
		\end{tabular}
		\vspace{-14 pt}
	\end{table}

	The deviation of voltage magnitude from the expected trend at bus 5 and 6 is consistent with other scenarios. The main reason behind that deviation is the active power consumption by the slack generator at bus 1 after a certain level of PV penetration. Until 50\% penetration, with the increment of PV integration (and the decrease of real power consumption of substation) the voltage rises both at bus 5 and 6. Here, the overall real power generated by 3 generators is balanced with the consumption of real power at the substations. But once the PV penetration goes higher than 50\%, the active power demand by the distribution systems is lower than the transmission system generation and hence the slack bus starts to consume active power. When the slack bus consumes the real power, the direction of power flow from bus 4 to bus 5 and bus 4 to bus 6 is reversed as shown in Figs. 5 and 6. Now the slack bus serves as a motor. Due to this the overall load of the system has not decreased.
	
	\begin{figure*}[t]
	\minipage{0.43\textwidth}
	\includegraphics[width=\linewidth, trim={0 0 0 0},clip]{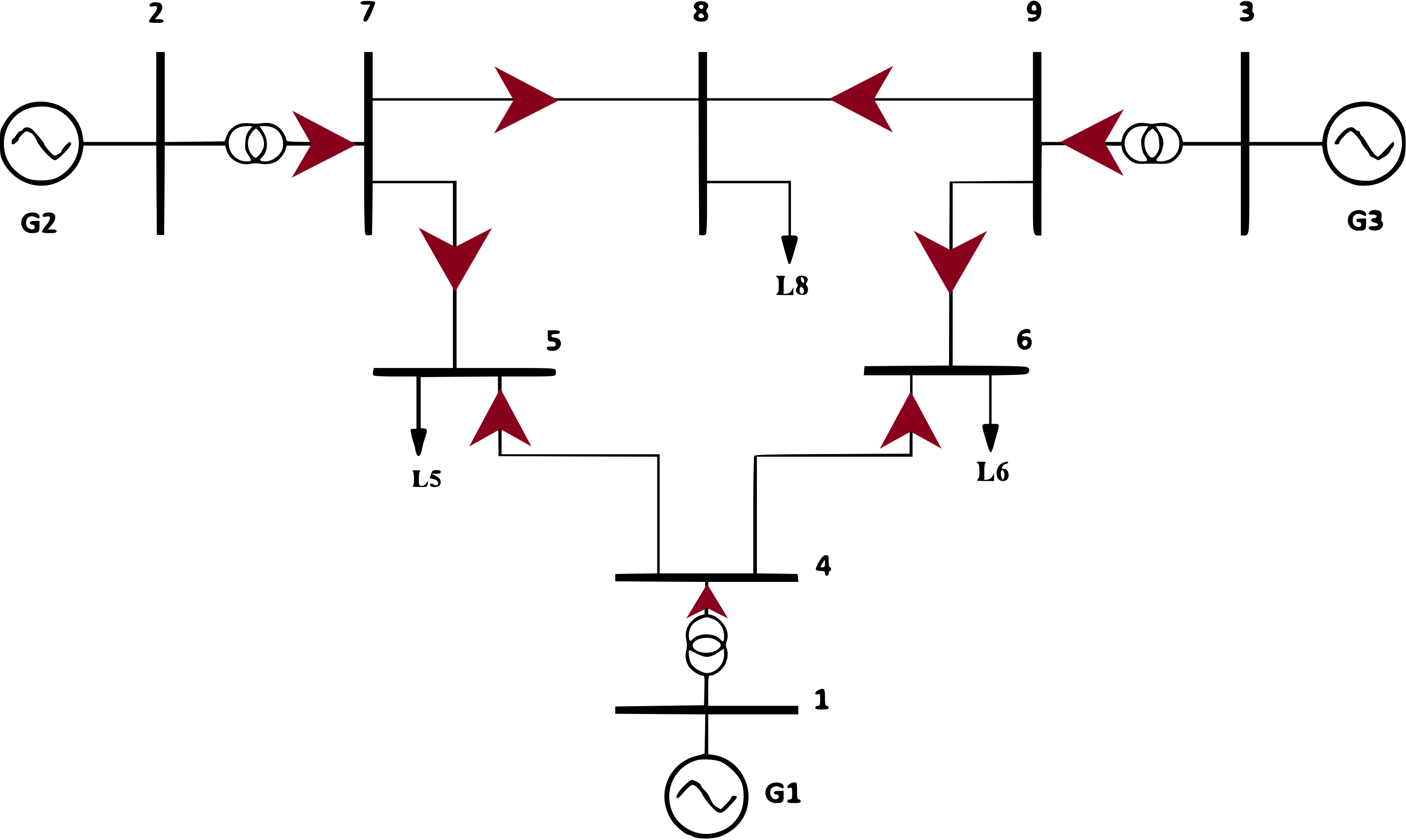}
	\caption{ Power flow direction until 50\% PV penetration}
	\endminipage\hfill
	\minipage{0.43\textwidth}
	\includegraphics[width=\linewidth,trim={0 0 0 0},clip]{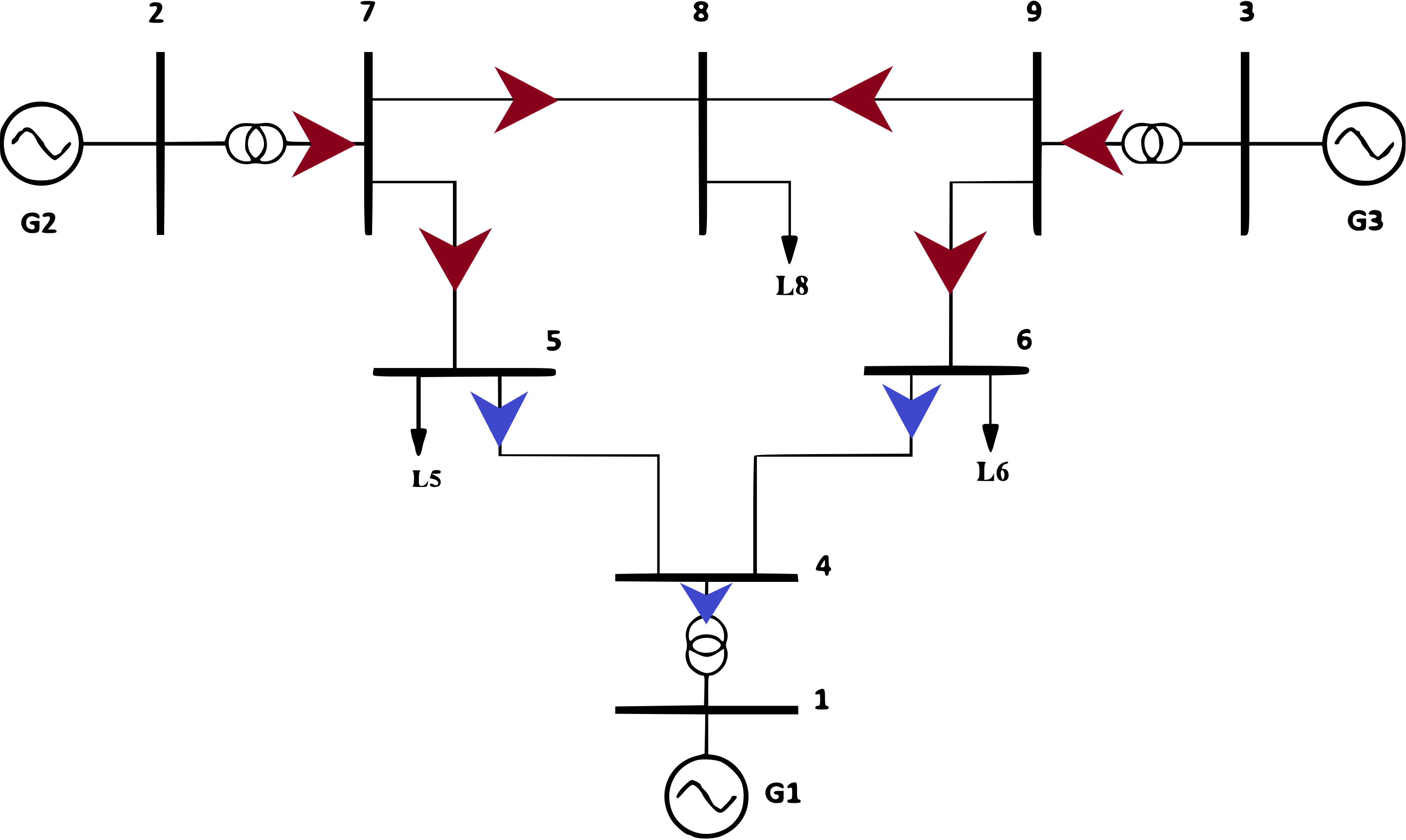}
	\caption{ Power flow direction after 50\% PV penetration}
	\endminipage\hfill
	\vspace{-0.5cm}
\end{figure*}

    With the increment of PVs  and subsequently the decrement of real power consumption at substations, slack bus absorbs additional real power from the system to balance the supply and demand. This happens, because the generation at bus 2 and bus 3 is fixed on account of being PV buses. The slack at bus 1 is connected to bus 5 and bus 6 via bus 4. So, when the real power consumption in L5, L6 and L8 decreases, slack bus consumes more power. This additional power flows through bus 5 and bus 6.  Even though the load demand at bus 5 and bus 6 decreases, due to the increased consumption of active power by the slack bus through bus 5 and 6, the overall real power consumption at those buses increases. Thus, beyond 50\% PV penetration, although the real power consumption reduces at the distribution substations, the bus voltages decreases on account of slack bus acting as a load to balance the supply and demand. 
    
    The increase of PV penetration has not significantly increased the unbalance in voltage magnitude. This is because the PV penetration cases are randomly generated without intentional unbalance. The unbalance at PCCs are not more than 0.2\%. In addition, the average number of iteration and time taken to converge for each case is presented in Fig 7.
	
	
		\begin{figure}[ht]
		\centering
		\includegraphics[width=0.4\textwidth]{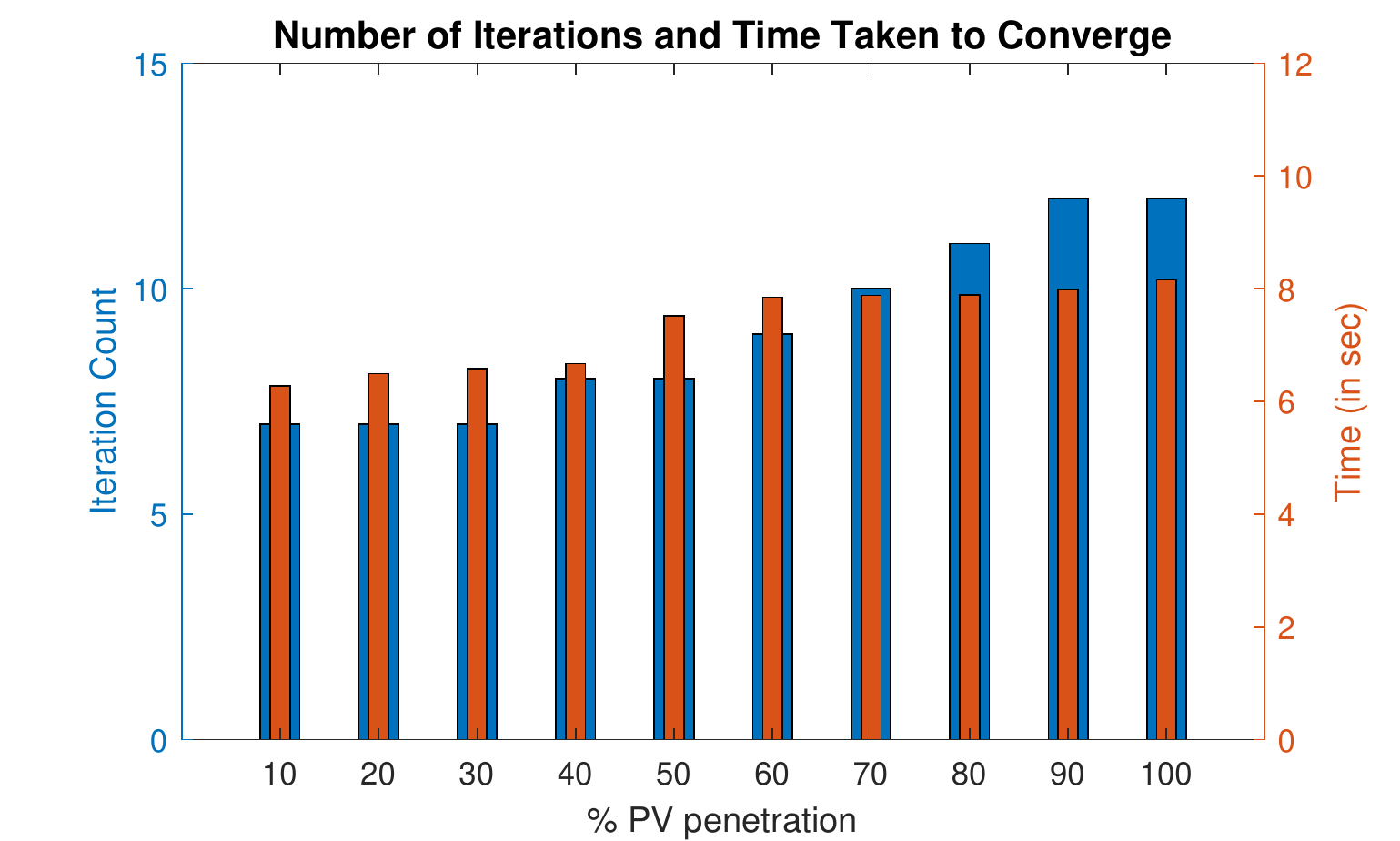}
		\vspace{-6 pt}
		\caption{Average number of iterations and time taken to converge by varying  PV penetration levels in 100 scenarios}
		\label{fig:5}
		\vspace{-6 pt}
	\end{figure}
	

	\subsection{Stand-alone model validation}
	The accuracy of the co-simulation model with increasing PV penetration is validated against stand-alone T\&D model developed using OpenDSS. Note that it also important to validate the voltage behavior observed at PCC in the previous Section. For the validation, scenario 25 is selected. Here, the co-simulation framework is referred as Model-1, and the standalone T\&D model is referred as Model-2. The voltages at PCC are compared in Table II for varying PV penetration levels. As can be observed from Table II, the voltages at bus 5 and bus 6 obtained by solving stand-alone model (Model-2) seems to follow the same trend as depicted in section III-A obtained using co-simulation (Model-1). The maximum difference of voltage between two models is less than 0.001 pu. 

	\vspace{-0.4 cm}
	\begin{table}[ht]
		\centering
		\caption{Converged positive-sequence voltages at T\&D PCC for varying PV penetrations using Model-1 and Model-2 }
		\label{singletable}
		\vspace{-0.2cm}
		\begin{tabular}{c|c|c|c|c|c|c}	
			\hline
			\multirow{2}{*}{ \% PV} &\multicolumn{3}{c|}{Model-1 Voltages (p.u.)}&\multicolumn{3}{c}{Model-2 Voltages (p.u.)} \\
			\cline{2-7}
			& Bus 5 & Bus 6 & Bus 8 & Bus 5 & Bus 6 & Bus 8 \\
			\hline
			10\%	&	1.0444	&	1.0511	&	1.0558	&	1.0444	&	1.051	&	1.0557	\\
			\hline
			20\%	&	1.0447	&	1.0516	&	1.0576	&	1.0446	&	1.0514	&	1.0574	\\
			\hline
			30\%	&	1.0449	&	1.0519	&	1.0588	&	1.0449	&	1.0518	&	1.0585	\\
			\hline
			40\%	&	1.0449	&	1.0521	&	1.0598	&	1.0449	&	1.0521	&	1.0595	\\
			\hline
			50\%	&	1.0448	&	1.0523	&	1.0608	&	1.0448	&	1.0522	&	1.0606	\\
			\hline
			60\%	&	1.0446	&	1.0522	&	1.0617	&	1.0446	&	1.0522	&	1.0616	\\
			\hline
			70\%	&	1.0442	&	1.0521	&	1.0625	&	1.0441	&	1.052	&	1.0627	\\
			\hline
			80\%	&	1.0438	&	1.0518	&	1.0633	&	1.0437	&	1.0517	&	1.0635	\\
			\hline
			90\%	&	1.0433	&	1.0514	&	1.0639	&	1.043	&	1.0513	&	1.0640	\\
			\hline
			100\%	&	1.0426	&	1.0509	&	1.0644	&	1.0424	&	1.0508	&	1.0646	\\
			\hline
		\end{tabular}
		\vspace{-0.2 cm}
	\end{table}
	
%

	\section{Conclusion}
	
	This paper employs a co-simulation platform to analyze the impacts of distribution-connected PVs on transmission system voltages. This co-simulation framework is iteratively coupled enabling an accurate simulation of integrated T\&D systems with results comparable to stand-alone unified T\&D simulation models.
	Multiple PV deployment scenarios based on locations and sizes of PV in the feeder are created and their impacts on transmission system voltages are studied. The study showed that the transmission voltages at the PCC decreases beyond 50\% PV penetration on account of slack bus acting as a load to balance the supply and demand. This effect on transmission voltages would have been difficult to analyze without co-simulating T\&D systems. Furthermore, the results obtained from this co-simulation model are thoroughly validated using a stand-alone T\&D system model.

	
	\bibliographystyle{ieeetr}
	\bibliography{references}
	
\end{document}